\documentclass[twocolumn,aps,preprintnumbers,aps]{revtex4-1}

\usepackage{graphicx}
\usepackage{dcolumn}
\usepackage{bm}

\begin{document}

\title{Energy loss by radiation to all orders in 1/N}

\author{Bartomeu Fiol and Blai Garolera}

\affiliation{Departament de F{\'\i}sica Fonamental i \\Institut de Ci{\`e}ncies del Cosmos, 
Universitat de Barcelona,\\
Mart{\'\i}\ i Franqu{\`e}s 1, 08028 Barcelona, Catalonia, Spain}

\email{bfiol@ub.edu, bgarolera@ffn.ub.es}

\begin{abstract}

We use the AdS/CFT correspondence to compute the energy radiated by an infinitely massive half-BPS particle charged under ${\cal N}=4$ $SU(N)$ SYM, transforming in the symmetric or antisymmetric representation of the gauge group, and moving in the vacuum, to all orders in $1/N$ and for large 't Hooft coupling. For the antisymmetric case we consider D5-branes reaching the boundary of $AdS_5$ at arbitrary timelike trajectories, while for the symmetric case, we consider a D3-brane in $AdS_5$ that reaches the boundary at a hyperbola. This D3-brane solution is the analytic continuation of the one previously used to compute the expectation value of a circular Wilson loop \cite{Drukker:2005kx}. We compare our results to the one obtained for the fundamental representation by Mikhailov \cite{Mikhailov:2003er}, obtained by considering a string in $AdS_5$.

\end{abstract}

\maketitle

\section{Introduction}
Given a gauge theory, one of the basic questions one can address is the energy loss of a particle charged under such gauge fields, as it follows arbitrary trajectories. For classical electrodynamics this is a settled question, with many practical applications \cite{jackson}. Much less is known for generic quantum field theories, especially in strongly coupled regimes. This state of affairs has started to improve with the advent of the AdS/CFT correspondence \cite{Maldacena:1997re}, which has allowed us to explore the strongly coupled regime of a variety of field theories. Within this framework, the particular question of the energy radiated by a particle charged under a strongly coupled gauge theory - either moving in a medium, or in the vacuum with non-constant velocity - has received a lot of attention (see \cite{Gubser:2009sn} for relevant reviews). The motivations are manifold, from the more phenomenological ones, like modeling the energy loss of quarks in the quark-gluon plasma \cite{Herzog:2006gh} to the more formal ones, like the study of the Unruh effect \cite{Chernicoff:2010yv}. In most of these studies the heavy particle transforms in the fundamental representation of the gauge group, and the dual computation is in terms of a string moving in an asymptotically AdS space. The main purpose of this note is to extend this prescription to other representations of the gauge group, which will amount to replace the fundamental string by D3 and D5 branes (see \cite{Chernicoff:2006yp} for a previous appearance of this idea), in complete analogy to the prescription developed for the computation of Wilson loops \cite{Rey:1998ik, Drukker:2005kx, Yamaguchi:2006tq, Gomis:2006sb,Hartnoll:2006is}.

Besides the intrinsic interest of this generalization, our main motivation in studying it is that, as it happens in the computation of certain Wilson loops, the results for the energy loss obtained with D-branes give an all-orders series in $1/N$. Given the paucity of such results for large $N$ 4d gauge theories, this by itself justifies its consideration. Furthermore, these $1/N$ terms might shed some light on some recent results in the study of radiation using the AdS/CFT correspondence. Let's briefly review them.

The case of an infinitely massive particle transforming in the fundamental representation and following an {\it arbitrary} timelike trajectory was addressed by Mikhailov \cite{Mikhailov:2003er}, who quite remarkably found a string solution in $AdS_5$ that solves the Nambu-Goto equations of motion and reaches the boundary at any given particle world-line. Working in Poincar\'e coordinates, 
\begin{equation}
ds_{AdS_5}^2=\frac{L^2}{y^2}\left(dy^2+\eta_{\mu \nu}dx^\mu dx^\nu\right)
\label{poincare}
\end{equation}
it was furthermore shown that the energy of that string with respect to the Poincar\'e time is given by 
\begin{equation}
E=\frac{\sqrt{\lambda}}{2\pi}\left(\int dt \frac{\vec a^2-|\vec a\wedge\vec v|^2}{(1-v^2)^3}+\gamma \frac{1}{y}|_{y=0}\right)
\label{mikha}
\end{equation}
where the integral is with respect to the world-line time coordinate, and $\lambda=g_{YM}^2 N$ is the 't Hooft coupling. The second (divergent) term corresponds to the (infinite) mass and $\gamma$ is the Lorentz factor. The first term corresponds to the radiated energy, so in the supergravity regime the total radiated power by a particle in the fundamental representation is
\begin{equation}
P_F=\frac{\sqrt{\lambda}}{2\pi}a^\mu a_\mu
\label{powerfun}
\end{equation}
which is essentially Lienard's formula for radiation in classical electrodynamics \cite{jackson} with the substitution $e^2\rightarrow 3\sqrt{\lambda}/4\pi$. This $\sqrt{\lambda}$ dependence also appears - and has the same origin - in the computation of the vev of Wilson loops at strong coupling \cite{Rey:1998ik, Maldacena:1998im}.

Having computed the total radiated power, a more refined question is to determine its angular distribution. For a particle moving in the vacuum, this has been done in \cite{Athanasiou:2010pv, Hatta:2010dz}, who found that this angular distribution is essentially like that of classical electrodynamics. This is a somewhat counterintuitive result, as one might have expected that the strong coupling of the gauge fields would tend to broaden the radiating pulses and make radiation more isotropic. In particular, the authors of \cite{Hatta:2010dz} argue that these results are an artifact of the supergravity approximation, and might go away once stringy effects are taken into account (see \cite{Hubeny:2010bq} for alternative interpretations). Here is where considering particles in other representations might be illuminating, since the $1/N$ expansion of the radiated power we find can be interpreted as capturing string loop corrections \cite{Drukker:2000rr}.

The plan of the present note is as follows: in the next section we introduce D5-branes dual to particles in the antisymmetric representation following arbitrary timelike trajectories, and evaluate the corresponding energy loss. We then consider a D3-brane dual to a particle in the symmetric representation following hyperbolic motion, and compute its energy loss. We end by discussing the possible connection of this result with the similar one for particles in the fundamental representation, and mentioning possible extensions of this work.

\section{D5 branes and the antisymmetric representation}
Given a string world-sheet that solves the Nambu-Goto action in an arbitrary manifold $M$, there is a quite general construction due to Hartnoll \cite{Hartnoll:2006ib} that provides a solution for the D5-brane action in $M\times S^5$, of the form $\Sigma \times S^4$ where $\Sigma \hookrightarrow M$ is the string world-sheet and $S^4\hookrightarrow S^5$. The evaluation of the respective renormalized actions gives then a universal relation between the vev of  Wilson loops in the antisymmetric and fundamental representations, already observed in particular examples \cite{Hartnoll:2006hr,Yamaguchi:2006tq}. More recently, this construction has been used to evaluate the energy loss of a particle in the antisymmetric representation, moving with constant speed in a thermal medium \cite{Chernicoff:2006yp}. In this section we combine Mikhailov's string world-sheet solution  \cite{Mikhailov:2003er} with Hartnoll's D5-brane construction \cite{Hartnoll:2006ib} to compute the radiated power for a particle in the antisymmetric representation.

For a given timelike trajectory, we consider a $D5$-brane in $AdS_5\times S^5$, with world-volume $\Sigma\times S^4$ where $\Sigma$ is the
corresponding Mikhailov world-sheet \cite{Mikhailov:2003er}. On $\Sigma$ there is in addition an electric DBI field strength with $k$ units of charge \cite{Hartnoll:2006ib}. This D5-brane is identified as the dual to a particle transforming in the k-th antisymmetric representation, and following the given timelike trajectory. As shown in \cite{Hartnoll:2006ib} the equations of motion force the angle of $S^4$ in $S^5$ to be
\begin{equation}
\sin \theta_0 \cos \theta_0 -\theta_0=\pi\left(\frac{k}{N}-1\right)
\label{anglekn}
\end{equation}
We now proceed to compute the energy with respect to the Poincar\'e time coordinate and the radiated power of such particle. The energy density for the D5-brane is
$$
{\cal E}_{D5}=T_{D5}\frac{L^2}{y^2}\frac{|\gamma+F|_s}{\sqrt{-|\gamma+F|}}=T_{D5}\frac{L^2}{y^2}\frac{|\gamma_\Sigma|_s}{\sin \theta_0 \sqrt{-|\gamma_\Sigma|}}\sqrt{|\gamma_{S^4}|}
$$
where the subscript $s$ means that the determinant is restricted to the spatial directions of the D5-brane or the fundamental string. We have used that in Hartnoll's solution the DBI field strength is purely electric and the DBI determinant is block diagonal. Integrating over the $S^4$ part of the world-volume one immediately obtains up to constants the energy density of the fundamental string, so
$$
E_{D5}=\frac{2N}{3\pi}\sin^3 \theta_0 E_{F1} 
$$
This is the same relation as the one found between the renormalized actions of the D5-brane and the fundamental string \cite{Hartnoll:2006ib}, and in \cite{Chernicoff:2006yp} for the relation of drag forces in a thermal medium. In the regime of validity of supergravity, the radiated power of a particle in the k-th antisymmetric representation is therefore related to the radiated power of a particle in the fundamental representation (\ref{powerfun}) by
\begin{equation}
P_{A_k}=\frac{2N}{3\pi}\sin^3 \theta_0 P_{F}
\label{poweranti}
\end{equation}

\section{D3 branes and the symmetric representation}
The computation of Wilson loops of half-BPS particles in the symmetric representation is given by evaluating the renormalized action of D3 branes \cite{Gomis:2006sb}, and analogously we propose to compute the radiated power of a half-BPS particle in the symmetric representation by evaluating the energy of a D3 brane that reaches the boundary of $AdS$ at the given timelike trajectory. Contrary to what happens for the fundamental or the antisymmetric representations, we currently don't have the generic D3-brane solution, so we will focus on a particular trajectory. On the other hand, since these D3-branes are fully embedded in $AdS_5$, we don't use any possible transverse dimensions, so the results should be valid for other 4d conformal theories with a gravity dual. 

The particular trajectory we will consider is one-dimensional motion with constant proper acceleration, which in an inertial system corresponds to $\gamma^3 a=1/R$. The trajectory is hyperbolic, $-(x^0)^2+(x^1)^2=R^2$. A relevant feature is that a special conformal transformation applied to a straight world-line (static particle) gives the two branches of hyperbolic motion \cite{rohrlich}. Besides its prominent role in the study of radiation and the Unruh effect, another reason to choose this trajectory is that the relevant D3-brane is the analytic continuation of the one already found in \cite{Drukker:2005kx}.

The radiated energy of a particle in the fundamental representation, eq.(\ref{mikha}) derived in \cite{Mikhailov:2003er}, is written in terms of the world-line of the heavy particle. At least in particular cases, it is possible to obtain an alternative derivation that emphasizes the presence of a horizon in the world-sheet metric, which encodes the split between radiative and non-radiative gluonic fields, and therefore signals the existence of energy loss of the dual particle, even in the vacuum \cite{Dominguez:2008vd}. It is convenient to briefly rederive this result for the particular case of hyperbolic motion, since the computation of the energy loss using a D3 brane that we will shortly present resembles closely this second derivation. Working in Poincar\'e coordinates, Mikhailov's string solution for hyperbolic motion can be rewritten as $y^2=R^2+(x^0)^2-(x^1)^2$; the Euclidean continuation of this world-sheet is the one originally used to evaluate the vev of a circular Wilson loop \cite{Berenstein:1998ij} (see also \cite{Xiao:2008nr}). This world-sheet is locally $AdS_2$ and has a horizon at $y=R$, with temperature $T=1/2\pi R$, which is the Unruh temperature measured by an observer following a $r_1=R$ trajectory in Rindler space. By integrating the energy density from the horizon to the boundary we obtain,
\begin{equation}
E=\frac{\sqrt{\lambda}}{2\pi}\int_0^R\frac{dy}{y^2}\frac{1}{\sqrt{R^2+(x^0)^2-y^2}}=
\frac{\sqrt{\lambda}}{2\pi}\left(\frac{-x^0}{R^2}+\gamma \frac{1}{y}|_{y=0}\right)
\label{mikhahyp}
\end{equation}
The contribution from the boundary is just the (divergent) second term, corresponding to the mass of the particle. The first term comes from the horizon contribution, and corresponds to the radiated energy.

\subsection{The D3-brane solution}
We are interested in a D3-brane that reaches the boundary of $AdS_5$ at a single branch of the hyperbola $-(x^0)^2+(x^1)^2=R^2$. To find it, we change coordinates on the $(x^0,x^1)$ plane of (\ref{poincare}) to Rindler coordinates, so the new coordinates cover only a Rindler wedge
\begin{equation}
ds^2=\frac{L^2}{y^2}\left(dy^2+dr_1^2-r_1^2d\psi^2+dr_2^2+r^2_2 d\phi^2\right)
\label{rindler}
\end{equation}
In these coordinates the relevant D3-brane solution found in \cite{Drukker:2005kx} is given by
\begin{equation}
(r_1^2+r_2^2+y^2-R^2)^2+4R^2r_2^2=4\kappa^2 R^2 y^2
\end{equation}
where
$$
\kappa=\frac{k\sqrt{\lambda}}{4N}
$$
Near the $AdS_5$ boundary $y=0$, this solution goes to $r_2=0, r_1^2=R^2$, so it reaches a circle in Euclidean signature and the branch of a hyperbola in the Lorentzian one. The D3-brane also supports a non-trivial Born-Infeld field-strength on its world-volume \cite{Drukker:2005kx}. By a suitable change of coordinates, its world-volume metric can be written as \cite{Drukker:2005kx}
\begin{equation}
ds^2=L^2(1+\kappa^2)(d\zeta^2-\hbox{sinh}^2 \zeta d\psi^2)+L^2 \kappa^2 (d\theta^2+\sin^2 \theta d\phi^2)
\end{equation}
so it is locally $AdS_2 \times S^2$, with radii $L\sqrt{1+\kappa^2}$ and $L\kappa$ respectively, and it has a horizon at $\zeta =0$ (i.e. $r_1=0$ in the coordinates of (\ref{rindler})). The temperature of this horizon can be computed by requiring that the associated Killing vector is properly normalized at infinity; this is easily done in the coordinates of (\ref{rindler}) and the resulting temperature is again
\begin{equation}
T=\frac{1}{2\pi R}
\end{equation}

\subsection{Evaluation of the energy}
To determine the total radiated power of this solution we will evaluate the energy with respect the Poincar\'e time coordinate $x^0$. The energy density is
\begin{equation}
{\cal E}=T_{D3}\left(\frac{L^2}{y^2}\frac{|\gamma +F|_s}{\sqrt{-|\gamma +F|}}-\frac{L^4}{y^4}\right)
\end{equation}
After we substitute the Lorentzian continuation of the solution of \cite{Drukker:2005kx} in this expression, the energy density is
\begin{equation}
{\cal E}=T_{D3}\frac{L^4}{y^4}\left(\frac{(1+\kappa^2)R^2+(x^0)^2}{\sqrt{\kappa^2(1+\kappa^2)R^4-\kappa^2 R^2 r_1^2-(1+\kappa^2)R^2r_2^2}}-1\right)
\end{equation}
The energy is the integral of this energy density from the boundary to the world-volume horizon. A long computation yields
\begin{equation}
E=\frac{2N\kappa}{\pi}\left(-\frac{x^0}{R^2}{\sqrt{1+\kappa^2}}+\gamma \frac{1}{y}|_{y=0}\right)
\end{equation}
Exactly as it happened for the string, eq. (\ref{mikhahyp}), the boundary contributes only the second term, which is divergent, and is just $k$ times the one for the fundamental string, eq. (\ref{mikhahyp}). The first term is the contribution from the horizon, and from it we can read off the total radiated power
$$
P_{S_k}=\frac{2N\kappa}{\pi}\sqrt{1+\kappa^2}\frac{1}{R^2}=\frac{k\sqrt{\lambda}}{2\pi}\sqrt{1+\frac{k^2\lambda}{16N^2}}\frac{1}{R^2}
$$
This result was found for a particular timelike trajectory with $a^\mu a_\mu=1/R^2$. Nevertheless, in classical electrodynamics the radiated power depends on the kinematics only through the square of the 4-acceleration, $a^\mu a_\mu$ and as we have seen, the same is true in theories with gravity duals for particles in the fundamental, eq. (\ref{powerfun}), and antisymmetric representations, eq. (\ref{poweranti}). It is then natural to conjecture that in the regime of validity of supergravity, the radiated power by a particle in the symmetric representation following {\it arbitrary} timelike motion is
\begin{equation}
P_{S_k}=\frac{k\sqrt{\lambda}}{2\pi}\sqrt{1+\frac{k^2\lambda}{16 N^2}}a^\mu a_\mu
\label{radsim}
\end{equation}
It would be interesting to check this conjecture by finding D3-branes that reach the $AdS$ boundary at arbitrary timelike trajectories and evaluating the corresponding energies.

We now discuss the range of validity of this result, and its possible relevance for the case of a particle in the fundamental ($k=1$) representation. By demanding that the radii of the D3 brane are much larger than $l_s$ and that its backreaction can be neglected, one can conclude \cite{Drukker:2005kx} that this result can be trusted when $N^2/\lambda^2\gg k \gg N/\lambda^{3/4}$. Is is therefore not justified a priori
to set $k=1$ in our result, eq. (\ref{radsim}). Nevertheless, the Euclidean continuation of this very same D3-brane was used in \cite{Drukker:2005kx} to compute the expectation value of a circular Wilson loop, which for $k=1$ is known exactly for all $N$ and $\lambda$ thanks to a matrix model computation \cite{Drukker:2000rr, Pestun:2007rz}, and it was found \cite{Drukker:2005kx} that the D3-brane reproduces the correct result in the large $N,\lambda$ limit with $\kappa$ fixed, i.e. even for $k=1$. This better than expected performance of the Euclidean counterpart of this D3-brane in a very similar computation suggests the exciting possibility that (\ref{radsim}) might capture correctly all the $1/N$ corrections to the radiated power of a particle in the {\it fundamental} representation, i.e. for $k=1$, in the limit of validity of supergravity.

An obvious question that we are currently pursuing is whether the angular distribution of the radiated energy obtained from this D3-brane differs qualitatively from the results obtained with fundamental strings \cite{Athanasiou:2010pv, Hatta:2010dz}.

Finally, as already mentioned, the Euclidean version of the D3-brane considered here was used in \cite{Drukker:2005kx} to evaluate the vev of a circular Wilson loop. That D3-brane result is in turn only an approximation to the {\it exact} result, available for all $N$ and $\lambda$ thanks to a matrix model computation \cite{Drukker:2000rr, Pestun:2007rz}. It would be extremely interesting to understand whether the radiated power of a particle coupled to a conformal gauge theory can be similarly computed by a matrix model.

\section{Acknowledgements} 
We would like to thank Mariano Chernicoff, Nadav Drukker, Roberto Emparan and David Mateos for helpful conversations. The research of BF is supported by a Ram\'{o}n y Cajal fellowship, and also by MEC FPA2009-20807-C02-02, CPAN CSD2007-00042, within the Consolider-Ingenio2010 program, and AGAUR 2009SGR00168. The research of BG is supported by an ICC scholarship and by MEC FPA2009-20807-C02-02.

\end{document}